\documentclass[10pt,letterpaper]{article}
\usepackage{opex3}
\usepackage{color}
\usepackage{amsmath}
\begin{document}

\title{Measuring the second order correlation function and the coherence time using random phase modulation }

\author{Chen-How Huang, Yung-Hsiang Wen and Yi-Wei Liu$^*$}

\address{Department of Physics, National Tsing Hua University, Hsinchu 30013, Taiwan }

\email{$^*$ywliu@phys.nthu.edu.tw} 



\begin{abstract}
A new approach to measure the second order correlation function $g^{(2)}$ and the coherence time was investigated. The $g^{(2)}$ was calculated from the photon pair time interval distribution by direct numerical self-convolution with the high order correction. The accuracy of this method was examined using an optical fiber based Hanbury-Brown-Twiss interferometer with a pseudo-thermal light source. We found that the significance of the high order correction is related to the factor $\bar{I}\tau_{c}$, which is the overlapping of the photon wave packets. A novel technique was also demonstrated to measure the coherence time $\tau_c$ of a light source using the random phase modulation. This method is more suitable for a weak light source with a long coherence time using a simple experimental setup.  
\end{abstract}


\section{Introduction}
\begin{figure}[h]
\begin{center}
\includegraphics[width=0.5\linewidth]{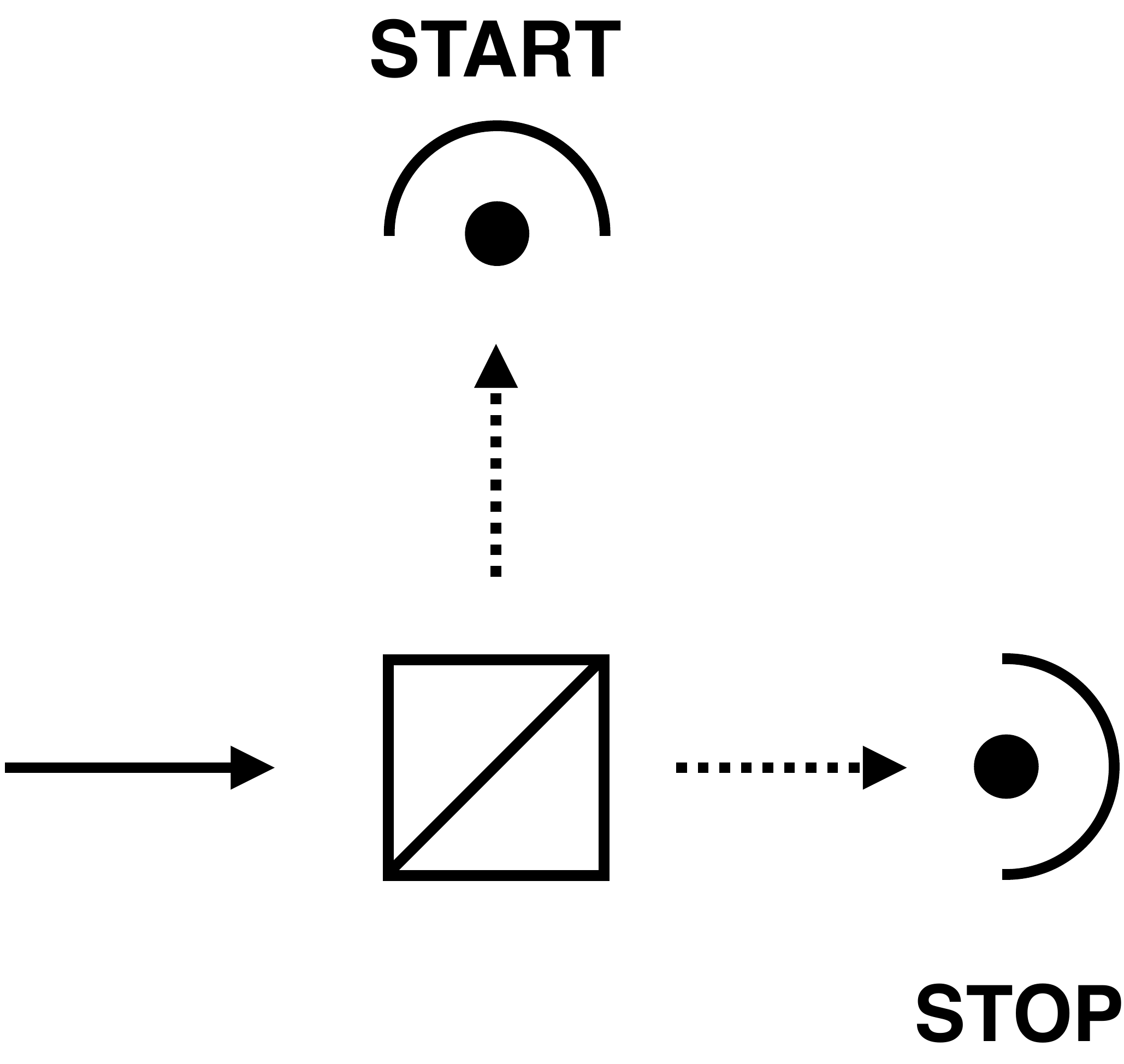}
\end{center}
\caption{\label{figure:simple}The simplified HBT experimental scheme. A clock (counter) is triggered by the photon received from the "START" detector, then is stopped by the subsequent photon received from the "STOP" detector. The time intervals were measured using the clock, and then recorded as a histogram. }
\end{figure}

The second order correlation function $g^{(2)}$ is one of the most important characteristic function for a light source~\cite{Plenio:1998ul}. It is the major feature to distinguish non-classical, anti-bunching light sources from the classical thermal ones. For a thermal light sources, the coherence time and the linewidth of the sources can also be directly derived from this function. To acquire an accurate $g^{(2)}$ plays an important role in various newly developed researches, such as quantum information \cite{NeergaardNielsen:2006hl}, fluorescence correlation spectroscopy on quantum dot \cite{Michler:2000wv}, cold atomic cloud \cite{Nakayama:2010vb, Das:2010vq}, single molecule \cite{DeMartini:1996vh,Fleury:2000uf} and et al. In the most of experiments, the Hanbury-Brown-Twiss (HBT) interferometer, whose simplified scheme is shown as Fig. \ref{figure:simple}, was implemented to measure $K(\tau)$, the probability distribution of photon pairs with time intervals $\tau$. The second order correlation function $g^{(2)}(\tau)$ can be approximated as $K(\tau)$, if the coherence times of the sources is relatively short and the photon flux is low. For the light sources with a very long coherence time, ex. fluorescence of ultracold atoms \cite{Du:2008do}, the direct conversion to $g^{(2)}(\tau)$ is problematic, due to the wave packet overlap of consecutive photons. Thus, reducing photon flux to avoid the overlapping could give a more accurate result of direct  $g^{(2)}$ conversion. However, the background noise level limits the achievable minimum photon flux. This dilemma constrains the applications of the method, which directly takes the photon pair time intervals as  $g^{(2)}(\tau)$. 

In this paper, we discuss the high order correction of the conversion to $g^{(2)}(\tau)$ from the photon pair time intervals $K(\tau)$ using the direct numerical convolution, and experimentally examined this method with a pseudo-thermal light source. A novel random phase modulation method is also demonstrated to measure the coherence time of a highly coherent source. It converts the source in the coherent sate to the chaotic state, in order to use the second order correlation function to characterize the (first order correlation) coherence time. Meanwhile, to overcome the dilemma of photon flux, as mentioned above, this method shortens the coherence time to allow a higher photon flux above the background noise.

To measure the second order correlation function using the HBT interferometer, two related physical quantities: $K(\tau)$ and $J(\tau)$ should be discussed as formulated in \cite{Reynaud:1983tq}. $K(\tau)$,  the experimentally measured result of the HBT,  is the histogram of the consecutive photon pairs with a time interval $\tau$. $J(\tau)$ is the histogram of photons at time=$\tau$ with a photon at t=0. The second order correlation function $g^{(2)}(\tau)$ is proportional to $J(\tau)$, as:
\begin{equation}
  J(\tau)=\bar{I}g^{(2)}(\tau)
\end{equation}
where $\bar{I}$ is the average photon detection rate per time bin of the light source and normalizes the histogram to the distribution of probability density. The time resolution (the bin size of the histogram) must be shorter than the coherence time of the light source under study. Since $J(\tau)$ should include the counts of the photon pair, which are not necessary to be consecutive, but just has a time interval $\tau$. $J(\tau)$ is an infinite convolution power series of $K(\tau)$  \cite{Reynaud:1983tq}, which is a histogram of the time intervals between two consecutive signal received from the "START" and the "STOP" detectors (see Fig. \ref{figure:simple}). Thus,
\begin{equation}{\label{eq:series}}
  J(\tau) = K(\tau)+K(\tau)*K(\tau)+......=\sum^{\infty}_{n=1}K_n(\tau)
 \end{equation}
 $K_n(\tau)$ is denoted as the self-convolution of $K(\tau)$ in $n$ orders. The Laplace transformed $\tilde{J}(s)$ then can be derived from $\tilde{K}(s)$ as:
\begin{equation}
  \tilde{J}(s) = \frac{\tilde{K}(s)}{1-\tilde{K}(s)}
\end{equation}
where $\tilde{J}(s)$ and $\tilde{K}(s)$ are the Laplace transforms of $J(\tau)$ and $K(\tau)$, respectively. Using this equation to calculate $g^{(2)}(\tau)$ requires an accurate and efficient numerical Laplace transformation and its inversion, which is very sensitive and a challenging task for numerical analysis. Therefore, it is not very often directly implemented in $g^{(2)}$ measurement experiments. In some of the experiments, it is to take $K(\tau)\sim J(\tau)$ as approximation by ignoring the high order terms, or to replace Laplace transformation with Fast-Fourier-Transformation (FFT) ~\cite{Fleury:2000uf,Fleury:2001uw}. In the following sections, we examined a direct numerical convolution algorithm to derive $g^{(2)}(\tau)$ from $K(\tau)$. The direct numerical convolution, which takes the high order correction into account,  can be computed using a simple recursion loop. The accuracy of this approach was tested using a pseudo-thermal light source, to show a significant improvement, particularly for a light source with a long coherence time.

\section{High-order correction of $g^{(2)}(\tau)$}{\label{sec:high-order}}

$J(\tau)$, proportional to $g^{(2)}(\tau)$, is a self-convolution power series of $K(\tau)$, as Eq.~(\ref{eq:series}). The order number of the self-convolution to reach a satisfactory accuracy depends on the convergency of the self-convolution power of $K(\tau)$. Conducting a Fourier transformation on Eq. (\ref{eq:series}) can simplify the successive convolution of $K(\tau)$ to the frequency domain $\hat{K}(\omega)$: 
\begin{equation}
\bar{I}\hat{g}^{(2)}(\omega)=\hat{J}(\omega)=\sum^{\infty}_{n=1}\hat{K}^{n}(\omega)=\frac{\hat{K}(\omega)}{1-\hat{K}(\omega)}
\end{equation}
It is valid, if $0< \left|\hat{K}(\omega)\right|<1$. A fast convergency will be given by a smaller $\hat{K}(\omega)$, the rate of the convergency can be quantified by :
\begin{equation}{\label{eq:k-omega}}
\hat{K}(\omega)=\frac{\bar{I}\hat{g}^{(2)}(\omega)}{1+\bar{I}\hat{g}^{(2)}(\omega)}
\end{equation}

For a chaotic thermal light source, we have
\begin{equation}\label{eq:1to2}
g^{(2)}(\tau) = 1+|g^{(1)}(\tau)|^{2}
\end{equation}
Then, for $\omega\neq0$, the Fourier transform of $K(\tau)$ is: 
\begin{equation}{\label{eq:k-f-ig}}
\hat{K}(\omega)=\frac{\bar{I}\|\hat{g}^{(1)}(\omega)\|^{2}}{1+\bar{I}\|\hat{g}^{(1)}(\omega)\|^{2}}
\end{equation}
The convergency is predominated by the product $ \bar{I}\|g^{(1)}(\omega)\|^{2}$, which is the power spectrum of the light source. 

A Lorentzian chaotic light source model was used to further investigate the convergency of $\hat{K}(\omega)$. The $g^{(1)}(\tau)$ of a homogenous broadened light source can be modeled as \cite{Loudor}:
\begin{equation}{\label{eq:l-model}}
g^{(1)}(\tau)=e^{-\frac{\tau}{\tau_{c}}}
\end{equation}
And, the power spectrum of such a light source is given as:
\begin{equation}{\label{eq:l-model-f}}
\|\hat{g}^{(1)}(\omega)\|^2=\frac{ \tau_c}{1+({\omega \tau_c/2})^2}
\end{equation}
By Eq. (\ref{eq:k-f-ig}),
\begin{equation}
\hat{K}(\omega)=\frac{\bar{I} \tau_c}{1+({\omega \tau_c}/{2})^2+{\bar{I} \tau_c}}
\end{equation}
It shows that $\hat{K}\sim0$ at the high frequency region, and the high order correction is only important in the low frequency region within the linewidth of the source, $\omega \tau_c<1$. Then, the convergency of the high order correction on calculating $g^{(2)}$ is dominated by the factor:   
\begin{equation}{\label{eq:c-factor}}
\frac{\bar{I}\tau_{c}}{\frac{5}{4}+\bar{I}\tau_{c}}<1
\end{equation}
The convergency of the power series $\sum\hat K^n(\omega)$ is related to the product of the average photon detection rate and the coherence time: $\bar{I}\tau_{c}$. A smaller $\bar{I}\tau_{c}$ can result in a faster convergency. The product $\bar{I}\tau_{c}$ can be used to characterize the degree of the overlapping of photon wave packets. For a stronger overlapping, the high order self-convolution will play a more important role. Thus, while the direct summation of finite high orders is utilized to calculate $g^{(2)}$, a lower photon flux $\bar{I}$ or a short coherence time $\tau_{c}$ needs to be satisfied for an accurate result.

The error of $J(\tau)$, due to only finite high order terms included in the numerical calculation, can be estimated as: 
\begin{equation}
\Delta \hat J_m(\omega)=\frac{\sum^{\infty}_{n=m+1}\hat K^{n}(\omega)}{\sum^{\infty}_{1}\hat K^{n}(\omega)}=\hat K^{m}(\omega)
\end{equation}
where $m$ is the highest order included in the finite power series. For a weak light source $\bar{I}\tau_c\ll1$, the error is $\sim(0.8\bar{I}\tau_c)^m$ and the convergency is with a power of $m$ . On the contrary, for the strong light sources with $\bar{I}\tau_c\gg1$, the error is approximated as $1-(1.25m/\bar{I}\tau_c)$, which goes down linearly with $m$. Therefore, it is important to have a sufficiently weak photon overlapping $\bar{I}\tau_c$ for a fast convergency. 

For a source with a long coherence time $\tau_c$, a very low photon flux rate $\bar{I}$ is required by an accurate calculation of the correlation function. However, although the photon flux of the source under the measurement can be reduced simply using an attenuator, the minimum photon rate is limited by the stray light or the dark counts of the electronic. The photon flux of the source must be significantly higher than the background noise to reach a reliable measurement. Therefore,  a dilemma is imposed on the measurement of $g^{(2)}$ of the light source with a long coherence time. In our following experiment, the rotating disk modulation method was demonstrated to overcome this obstacle. The coherence time of a long photon, a single frequency Ne-He laser, was measured. 

\section{Uncertainty of the Beam splitter ratio}
As our discussion above, it seems that the more high order correction are included, the accuracy of the $g^{(2)}$ will be higher. In this section, we argue that the uncertainty of the beam splitter ratio will limit the highest order that can be included in the calculation. 

To use the direct numerical convolution method, because of the partial reflection beam splitter, the splitting ratio correction factor must be taken into account. The experimentally measured time interval is not truly of two consecutive photons. In HBT measurement, after the "START" detector receiving a photon, the consecutive photon may go to the "START" detector, rather than the "STOP" detector, with a probability of 1/2 for a 50:50 beam splitter. In such a case, there will be no signal generated from this event. Thus, the time interval histogram of the consecutive photon pairs $K(\tau)$ should be related to the experimentally measured photon pair time interval distribution $D(\tau)$ as:
\begin{equation}
D(\tau)=\sum^{\infty}_{n=1}\frac{1}{2^n}K_n(\tau)
 \end{equation}
 The $m$th order self-convolution $D_m(\tau)$s are expressed as:
 \begin{equation}
  \begin{aligned}
  D(\tau)     =&\frac{1}{2}K(\tau)+&\frac{1}{4}K_2(\tau)+&\frac{1}{8}K_3(\tau)+&\frac{1}{16}K_4(\tau)\dots\\
  D_2(\tau)=&                &\frac{1}{4}K_2(\tau)+&\frac{2}{8}K_3(\tau)+&\frac{3}{16}K_4(\tau)\dots\\
  D_3(\tau)=&                &                  &\frac{1}{8}K_3(\tau)+&\frac{3}{16}K_4(\tau)\dots\\
  D_4(\tau)=&                &                   &                                 &\frac{1}{16}K_4(\tau)\dots\\
 &\vdots\\
 \end{aligned} 
 \end{equation}

 Thus, $J(\tau)$ can be calculated from the self-convolution power series of the experimentally measured time interval distribution $D(\tau)$ with an additional factor 2, while a 50:50 beam splitter was used. 
 \begin{equation}{\label{eq:D2J}} 
2\sum^{\infty}_{n=1}D_n(\tau)=2(\frac{1}{2})\sum^{\infty}_{n=1}K_n(\tau)=J(\tau)\\
 \end{equation}

However, practically, a non-equal splitting ratio should be considered. A deviation $\epsilon$, which could be caused by the imbalanced beam splitting ratio or the difference between the efficiencies of the detectors, leads to a correction of the 50\% detection probability ratio as $(0.5+\epsilon):(0.5-\epsilon)$. The experimentally measured $D(\tau)$ is then written as:
\begin{equation}
 D(\tau,\epsilon)=\sum^{\infty}_{n=1}(\frac{1}{2}+\epsilon)^{n-1}(\frac{1}{2}-\epsilon)K_n(\tau)
\end{equation}
The uncertainty of the splitting ratio $\epsilon$ will affect the accuracy of the resulted $g^{(2)}$. The higher order self-convolution terms will be more severely affected. That is, such uncertainties will be amplified in the high order terms and limits the final achievable accuracy. A deviation $\epsilon$=5\% in the splitting ratio, will result in a 30\% error for the 5th order, and 100\% for the 10th order. The experimental difficulty in measuring an accurate splitting ratio limits the maximum applicable order in calculating the final second order correlation function.    

\section{HBT experimental test}

\begin{figure}[h]
\begin{center}
\includegraphics[width=0.8\linewidth]{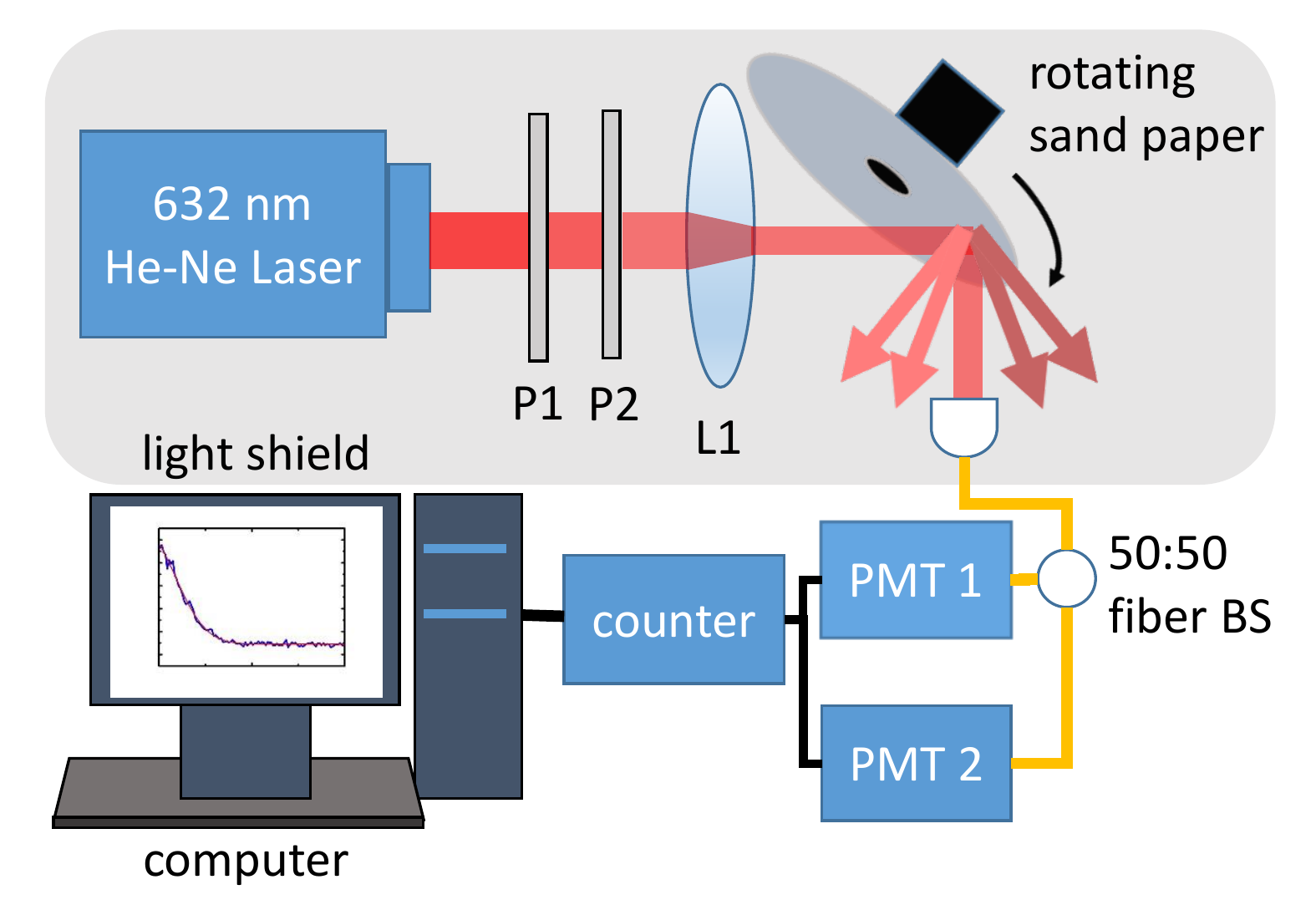}
\end{center}
\caption{\label{figure:setup}The experimental set-up. To generate a pseudo-thermal light source, a single longitudinal mode 632~nm He-Ne laser, passing through two polarizers for controlling the incident power, is focused on a rotating wheel with a surface of sandpaper. The back scattering of light was collected into a fiber splitter without any collimator.  One of SPCM (PMT1) was as the START to trigger the universal counter for time interval measurement. The other SPCM (PMT2) was the "STOP". The time intervals were recorded by a computer for subsequent off-line analysis. The second order correlation function of the pseudo-thermal light was then calculated from the histogram of the recoded time intervals. }
\end{figure}

The improvement and the accuracy of the direct numerical convolution method for analyzing $\rm g^{(2)}(\tau)$ was tested using a HBT photon interferometer to measure a pseudo-thermal light source with a variable coherence time. Meanwhile, to measure a very long coherence time (a narrow linewidth), we demonstrate a novel method, which is based on this experimental setup and needs no km-long optical fiber for self-heterodyne detection \cite{OKOSHI:1980un,Richter:1986us}.

As shown in Fig.\ref{figure:setup}, the HBT interferometer is based on a fiber-splitter (50:50, throlabs 780-HP) with single photon counting modules (SPCM, HAMAMATSU H7421-40) as the "START" and the "STOP" detectors to measure the photon pair time intervals using an universal counter (Agilent 53131A).  The pseudo-thermal light is the back scattering of the single frequency 632~nm He-Ne laser from a rotating sand disk wheel.  No frequency or power stabilization was utilized onto the He-Ne laser. It is particularly suitable to simulate a light source with a very low intensity and a very long coherence time \cite{Martienssen:1964ig, Scarcelli:2004bc, ESTES:1971}. The coherence time of the pseudo-thermal light was controlled by the rotating speed $\omega_r$ of the sand paper (as ${\tau_c\propto1/\omega_r}$) to allow us to compare the deduced coherence time with the theoretical predication. The He-Ne laser power was controlled by rotating two polarizers, then focused on the rotating sand disk using a convex lens. For a good stability and precision of the rotating speed of the sand disk, the rotating wheel was modified from a light chopper wheel, whose rotating speed was locked to an oscillator clock. 

Reducing the stray room light is important for measuring a long coherence time. As shown by Eq. (\ref{eq:c-factor}), the product of the average photon detection rate and the coherence time ${\bar{I} \tau_c}$ should be smaller than 1 for an accurate measurement. That is, the longer coherence time requires a lower photon rate. On the other hand, the detected photon counts of the light source should be much higher than the background photon counts. We have reinforced the light shield for the entire experimental setup, including the fiber jackets and the connectors. The resulted background photon counting rate is about 20\% of the pseudo-thermal light. 

\begin{figure}
\begin{center}
\includegraphics[width=0.8\linewidth]{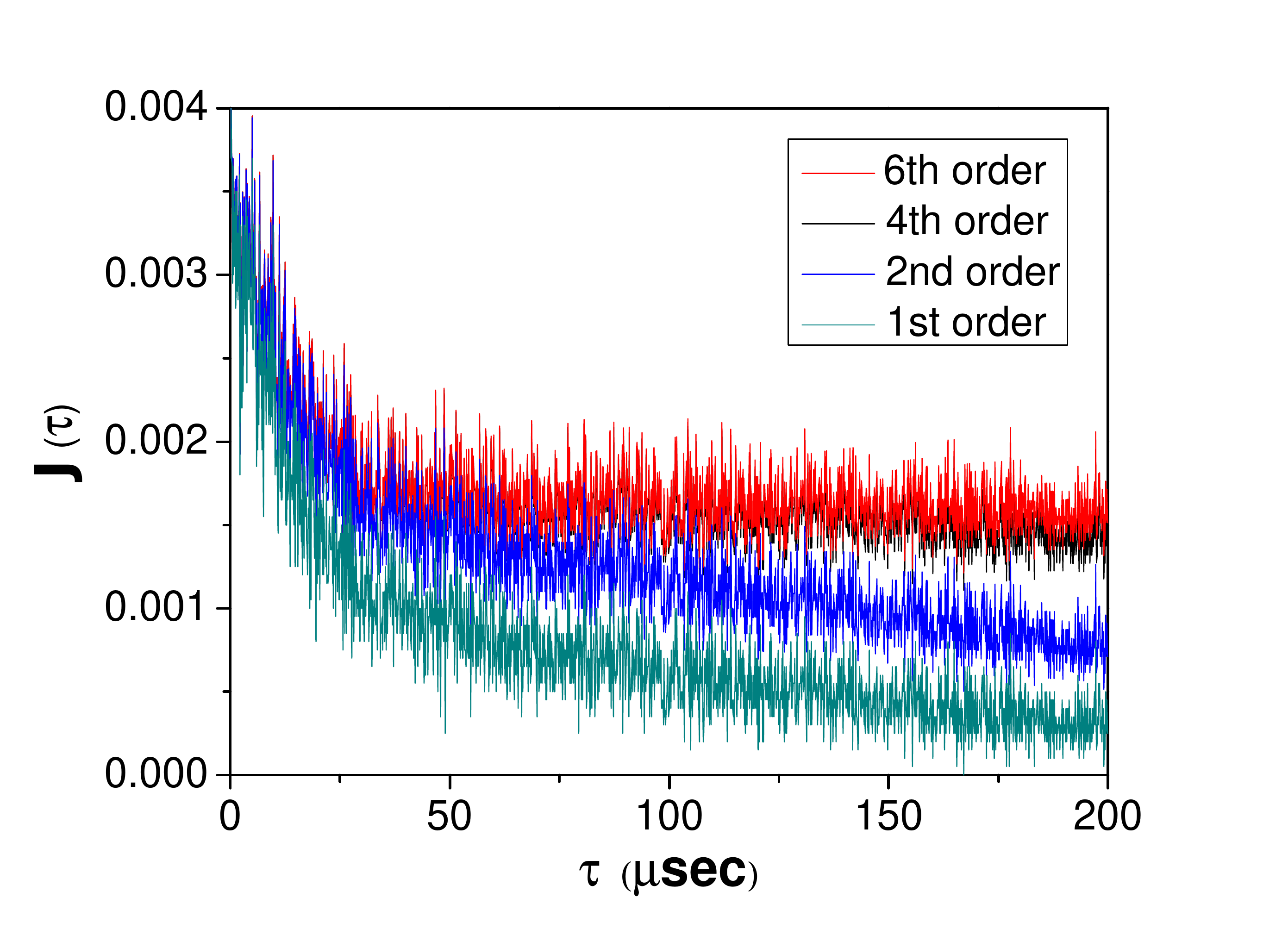}
\end{center}
\caption{\label{fig:contribution} Typical $J(\tau)$ with various orders of $D_n(\tau)$ at a 20~Hz rotating frequency. The coherence time of the pseudo-thermal light source $\tau_{c}$ is $\sim 10\mu sec$ and $\bar{I} \sim 4\times10^4$ photons/sec. The results of 4th and 6th order are in a very good agreement within the region $\tau<100~\mu sec$, where is important for deriving the coherence time. The bin size = 100~nsec was used. }
\end{figure}

In our experiment, the detected photon rate was reduced to $\sim4\times10^4$ photons/sec. A recorded time interval sequence was then converted to a histogram ($D(\tau)$) with a 100~nsec bin size of time, which is sufficiently small for the measurement of a coherence time $\sim\mu$sec. A smaller bin size has also been used for our test run and showed no improvement on the result, but just cost more calculation time.

$J(\tau)$, which is proportional to the second order correlation function $g^{(2)}(\tau)$, was derived using direct numerical self-convolution of $D(\tau)$ (Eq. (\ref{eq:D2J})). Figure \ref{fig:contribution} shows the $J(\tau)$s with various high order corrections. We found that a six-order convolution is sufficient to converge and to reach an accurate $J(\tau)$. Taking the distribution of the photon pair intervals as $g^{(2)}$ or underestimating the high order correction will result in a prolonged coherence time. It also shows that the convergency is faster at the region of small $\tau$ as discussed in the section \ref{sec:high-order}. On the contrary, the tail of $J(\tau)$ with a large $\tau$, which does not affect the resulted coherence time $\tau_c$, has a slower convergency. We suggest the proper criterion of the highest order $n$ included in the calculation to be: $D_{n+1}/\sum^{n}_{1}D_n<10^{-4}$ in the region of $\tau<\tau_c$, since the deviation of the resulted coherence time will be $<1\%$. In our experiment, $n=6$ is sufficient to meet the criterion.

\section{Novel method of measuring the coherence time}
\begin{figure}
\begin{center}
\includegraphics[width=0.8\linewidth]{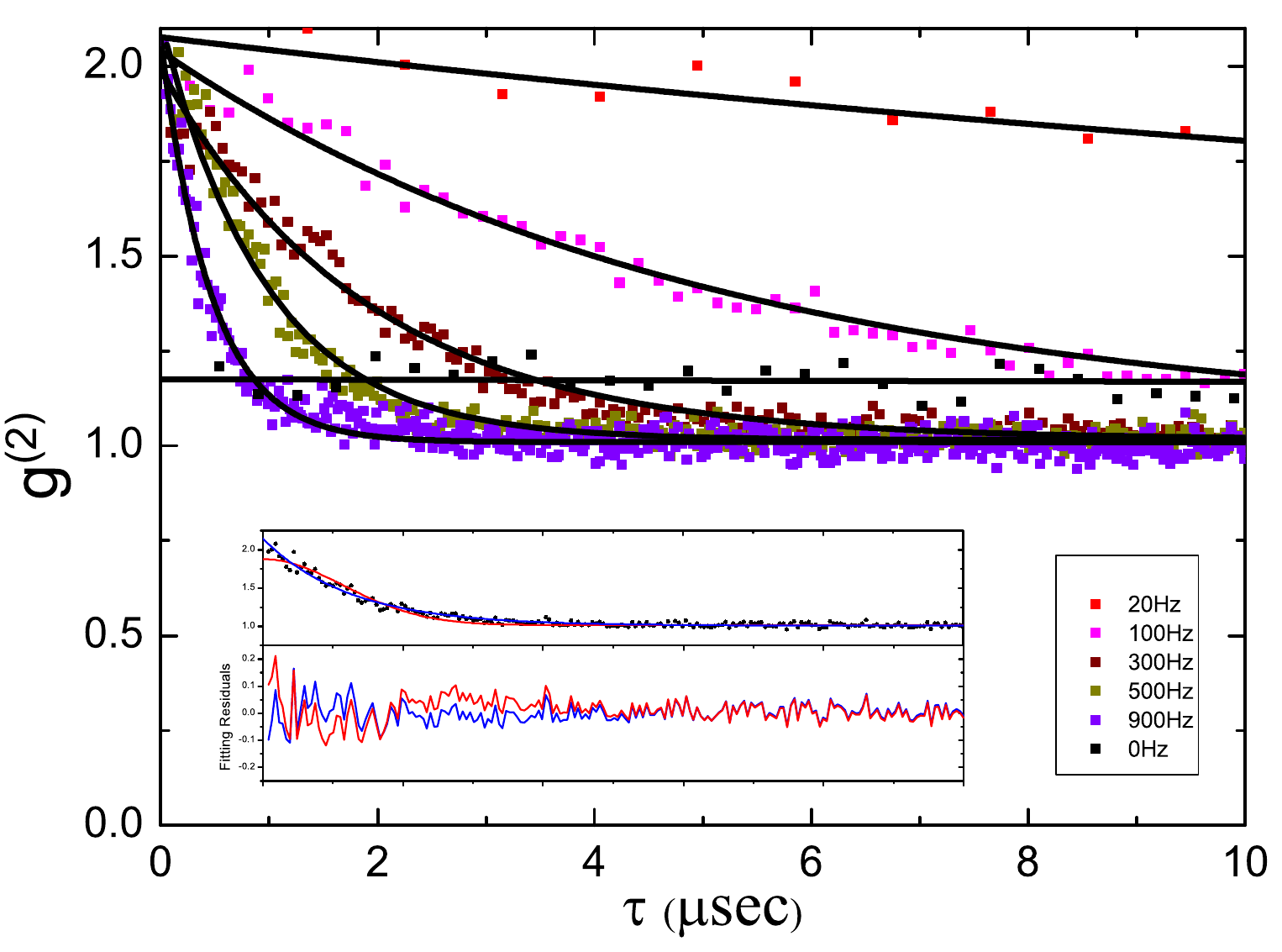}
\end{center}
\caption{\label{fig:fitting}The second order correlation functions $g^{(2)}(\tau)$ with rotating frequencies 0~Hz, 20~Hz, 100~Hz, 300~Hz, 500~Hz and 900~Hz. The black lines are the fitting functions $A+Be^{-2\tau/\tau_{c}}$. Except the 0~Hz, the resulted coherence times ${\tau_c}$ are $\rm{28.10(80)~\mu s}$, $\rm{7.40(11)~\mu s}$, $\rm{3.00(5)~\mu s}$, $\rm{1.75(2)~\mu s}$ and $\rm{0.96(3)~\mu s}$, respectively. The inset shows a Gaussian fit (thin red line) of the $g^{(2)}(\tau)$ with a rotating frequency 700 Hz, in comparison with the exponential decay fit (thin blue line). The fitting residual and $\chi^2$ show that the exponential decay function is slightly better than the Gaussian.}
\end{figure}

To test and demonstrate the accuracy of $g^{(2)}(\tau)$ ($J(\tau)$) using our method, the rotating speed of the sand disk was varied to generate photons with different coherence time in our experiment. The result is shown in Fig. \ref{fig:fitting}. The measured $J(\tau)$s using our direct convolution method were fitted with exponential decay curves with constant offsets, as: $A+Be^{-2\tau/\tau_{c}}$. From Eq. (\ref{eq:1to2}), we have \cite{Loudor}:
\begin{equation}
g^{(2)}(\tau)=1+e^{-2\tau/\tau_c}
\end{equation}
The coherence times of the scattered light (pseudo-thermal, chaotic) can be derived from the fitting parameter $\tau_c$. It is a direct relation between the second order and the first order correlations. However, this equation is only valid for a chaotic light source. For a coherent source, such as lasers, the coherence time can not be measured using the second order correlation. In our experiment, the coherent laser source was thus converted to the chaotic using the random phase modulation (rotating sand disk). In Fig.\ref{fig:fitting}, the zero time delay second order correlations $g^{(2)}(0)$ are about 2 for all the rotating frequencies, except for the 0~Hz. It indicates that the rotating sand disk has fully "thermalized" the coherent light to be chaotic, due to the random phase modulation. However, for the 0~Hz, the light was still in a very good coherent state, $g^{(2)}(0)\sim1$.        

The random phase modulation is not only to convert the light source to be pseudo-thermal, but also to broaden the linewidth. The broadened linewidth is proportional to the rotating speed. A composite (voigt) power spectrum with both Gaussian and Lorentzian parts could be the most adequate shape for our pseudo thermal light source. However, due to limited accuracy of the measurements, we are not able to determine the composition ratio, as \cite{Yuan:2009ue}. Both lineshape models are not much different to our experiment. As shown in the inset of Fig. \ref{fig:fitting}, the exponential decay curve (Lorentzian power spectrum) is more suitable than the Gaussian (Gaussian power spectrum). It is due to that the backscattering random modulation is similar to the mechanism of the collision broadening, which broadens the spectrum of light source as a Lorentzian shape. 

\begin{figure}
\begin{center}
\includegraphics[width=0.8\linewidth]{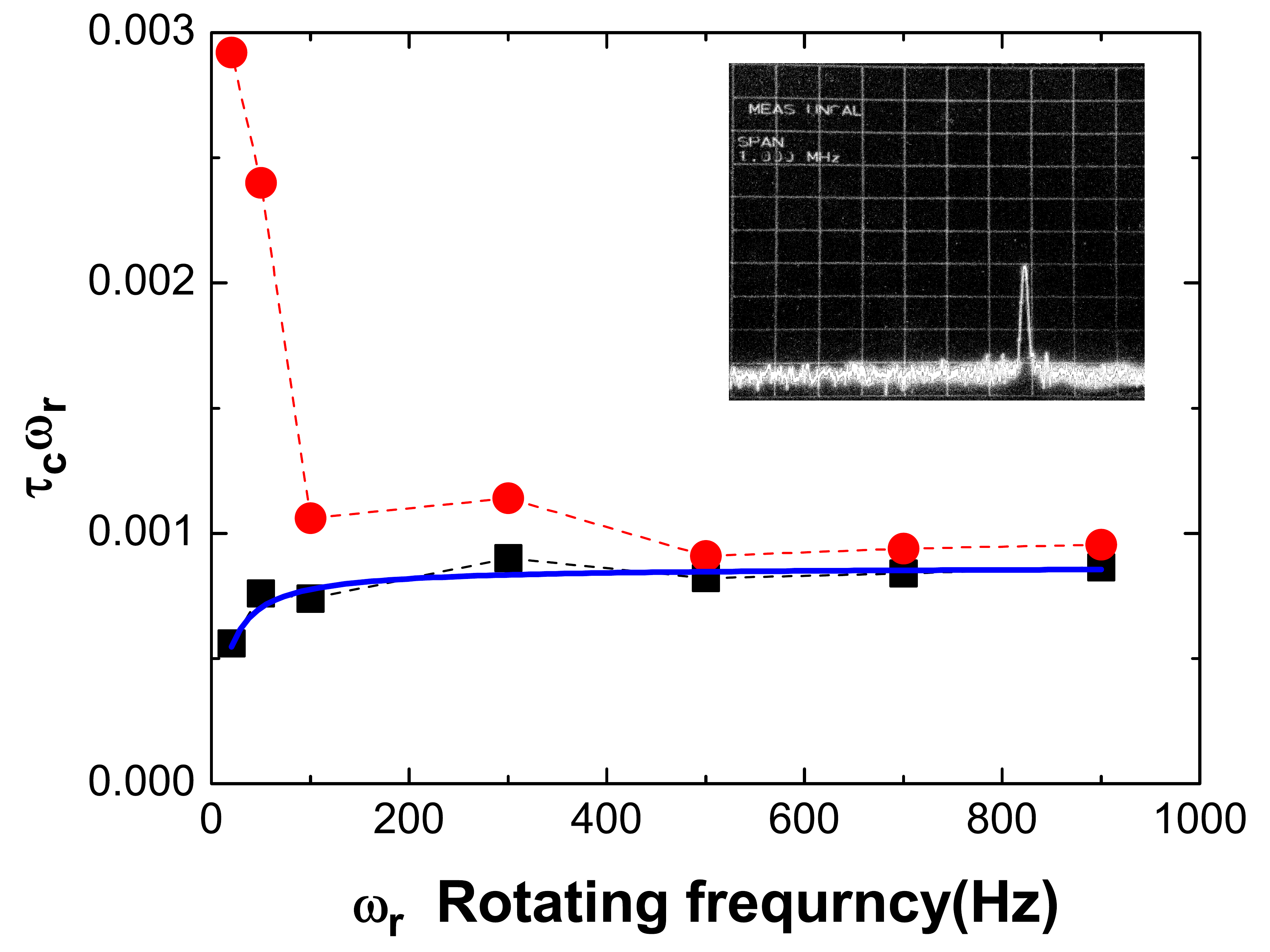}
\end{center}
\caption{\label{fig:comp} $\tau_{c}\omega_{r}$ v.s. $\omega_{r}$. At the low frequency regime, the uncorrected ${\tau_c}$ (red dot) strongly deviates from the high-order corrected $\rm{\tau_c}$ (black square) and the theoretical model (blue line). The corrected ${\tau_c}$ are in very good agreement with the theory, which gives $ \tau_{c}\omega_r=({(\omega_r \tau_{0}})^{-1}+k)^{-1}$. The coherence time ${\tau_0}$ of the incident light (He-Ne laser) was derived as 74(15)~$\mu s$ from the fitting parameter of the theoretical model. The inset shows a typical beat-note signal of two HeNe lasers with a RBW=3~kHz. The measured (-3db) linewidth is 6.5(1.3)~kHz. Assuming equal linewidth of the two lasers,  the coherence time is $\rm{97(20)~\mu s}$.}
\end{figure}

While the incident light is a monochromatic i.e. laser, with a negligible linewidth $\delta\omega_0\sim 0$, the coherence time of the pseudo-thermal light is proportional to the inverse of the rotating frequency \cite{ESTES:1971}.      
\begin{equation}
\frac{1}{\tau_{c}}=k\omega_r
\end{equation}
where k is a constant related to the experimental setup of the pseudo-thermal light. It is a good approximation for a high rotating frequency, whose scattering broadening is much larger than the laser linewidth itself. Thus, 
\begin{equation}
\tau_c\omega_r=\frac{1}{k}=\rm{const.} 
\end{equation}
can be used to test the validity of the derived coherence times $\tau_c$. As illustrated in Fig.~\ref{fig:comp}, $\tau_c\omega_r$ exhibits as a horizontal straight line at the higher rotating frequency $\omega_r$ regime.
However, for a finite linewidth $\delta \omega_o$ of the incident light, the resulted spectrum should be considered as the convolution of the incident light and the broadening caused by the rotating sand disk. The total linewidth $\delta\omega_m$ of the scattered pseudo-thermal light was then corrected as:
\begin{equation}
 \delta\omega_{m}= \delta\omega_{0}+k\omega_r\\
\end{equation}
And, the coherence time of the pseudo-thermal light, including the incident laser linewidth, is:
\begin{equation}
 \frac{1}{\tau_{c}}=\frac{1}{\tau_{0}}+k\omega_r\\
\end{equation}
where $\tau_0$ is the coherence time of the incident light. Consequently, $\tau_c\omega_r$ is no longer a constant, but a function of the rotating frequency $\omega_r$:
\begin{equation}\label{eq:model}
 \tau_{c}\omega_r=\frac{1}{\frac{1}{\omega_r \tau_{0}}+k}
 \end{equation}
It implies that the finite linewidth $\delta\omega_0$ correction becomes very pronounced at the low rotating frequency, where the scattering broadening $k\omega_r$ is comparable to the incident linewidth $\delta\omega_0$ (=$1/\tau_0$), and $ \tau_{c}\omega_r$ remains as a constant at the high rotating frequency regime, where $k\omega_r>>\delta\omega_0$.

Figure~\ref{fig:comp}, which shows $\tau_{c}\omega_r$ v.s. the rotating frequency $\omega_r$, is used to evaluate the validity of the high-order correction. Firstly, the high-order corrected data points (black) are in a very good agreement with the theoretical model Eq.~(\ref{eq:model}). The product $\tau_{c}\omega_r$ remains constant at the high frequency regime, and decreases at the low frequency due to the finite linewidth of the light source. In contrast, $\tau_{c}\omega_r$ of the uncorrected data points (red) increases at the low frequency and deviates further away from the theoretical prediction, because of the strong overlapping of the photon wave packets, i.e., a large $\bar I\tau_c$. Secondly, The high-order corrected data were fitted using the mathematical model $y=1/(A+B/x)$ (blue line). The finite coherence time of the unstabilized laser $\tau_0$ was derived from the parameter 1/B. 

The derived laser coherence time ${\tau_0}$ is 74(15)$\mu$s with a statistic uncertainty of 20\%, which was given by the fitting to the data points. The error-bars of each data points are given by Fig.~\ref{fig:fitting}, and too small to show in Fig.~\ref{fig:comp}. They are mainly caused by the low frequency noise as our discussion in section \ref{sec:high-order}. Generally, the uncertainty of the coherence time can be improved by more measurements at various rotating frequency, especially at the low rotating frequency regime, where $\omega_r\tau_0\sim1/k$. $k$ is a parameter related to the experimental setup, including the sand disk roughness, the distance between the scattering spot and the centre of rotation, and so on. However, for coherent incident light source (e.g., a laser), the lowest applicable rotating frequency must provide sufficiently strong random modulation to thermalize the source. In our experiment, a 20~Hz rotating frequency can fully thermalize the laser source ($g^{(2)}(0)=2$). A larger photon collecting aperture could maintain the thermalization with a lower rotating frequency \cite{Martienssen:1964ig}.

The frequency beat-note experiment with two HeNe lasers was also performed to measure the laser linewidth, and to compare with the results of the random phase modulation. The laser used for the random phase modulation measurement was mixed with another nearly identical HeNe laser using a beam splitter. The beat-note of these two lasers was measured using an avalanche-photodiode APD and a radio-frequency spectrum analyzer with a 3~kHz resolution-bandwidth (RBW). The beat-note signal is shown as the inset of Fig.~\ref{fig:comp}. The measured (-3db) linewidth of the beat-note is 6.5(1.3)~kHz.  Assuming that this width was equally contributed from two nearly identical lasers ($\Delta f$=3.3(0.7)~kHz for one laser), the coherence time $\tau_c$ ($=(\pi \Delta f)^{-1}$) is $\rm{97(20)~\mu s}$, which is in agreement with our random phase modulation measurement.

\section{Conclusions}
To measure the second order correlation function $g^{(2)}(\tau)$ of a light source, we have examined the feasibility and the reliability of the direct numerical convolution method, which is efficient and straightforward, in comparison with the other more delicate methods. The significance of the high order correction is related to the factor $\bar{I} \tau_c$, which indicates the overlapping between the wave packets of photons. It has been experimentally tested using a pseudo-thermal light source with a variable coherence time. In our experiment, we found that the summation up to the 6th order self-convolution can reach an accuracy sufficient to derive the second order correlation function $g^{(2)}$. The advantage of the direct numerical self-convolution is that it can be implemented with fast digital logic electronics, such as Field-programmable gate array (FPGA), to obtain a `real time' $g^{(2)}$ with a higher accuracy.

By applying this direct self-convolution method, a novel random phase modulation method of measuring the linewidth (coherence time) of a light source has been demonstrated. It is to use a rotating sand disk to randomly modulate the phase and broaden the linewidth of the light source, and then the linewidth (coherence time) of the source can be extrapolated to the zero modulation. In comparison with the commonly used self-heterodyne measurement, our method, which needs no kilo-meter long optical fiber and high intensity, is more favourable for a weak light source, such as molecule fluorescence, bio-photon emission et al., whose interference fringes or beat note are not easy to be detected.    

\section*{Acknowledgments}
This work was supported by the Ministry of Science and Technology of Taiwan under grant no. 103-2112-M-007-007-MY3.


\begin{thebibliography}{10}
\newcommand{\enquote}[1]{``#1''}

\bibitem{Plenio:1998ul}
M.~B. Plenio and P.~L. Knight, \enquote{{The quantum-jump approach to
  dissipative dynamics in quantum optics},} Rev Mod Phys  (1998).

\bibitem{NeergaardNielsen:2006hl}
J.~S. Neergaard-Nielsen, B.~M. Nielsen, C.~Hettich, K.~M{\o}lmer, and E.~S.
  Polzik, \enquote{{Generation of a Superposition of Odd Photon Number States
  for Quantum Information Networks},} Phys. Rev. Lett. \textbf{97}, 083604
  (2006).

\bibitem{Michler:2000wv}
P.~Michler, A.~Imamo{\u g}lu, M.~D. Mason, and P.~J. Carson, \enquote{{Quantum
  correlation among photons from a single quantum dot at room temperature},}
  Nature  (2000).

\bibitem{Nakayama:2010vb}
K.~Nakayama, Y.~Yoshikawa, H.~Matsumoto, Y.~Torii, and T.~Kuga,
  \enquote{{Precise intensity correlation measurement for atomic resonance
  fluorescence from optical molasses},} Opt Express \textbf{18}, 6604--6612
  (2010).

\bibitem{Das:2010vq}
M.~Das, A.~Shirasaki, K.~P. Nayak, M.~Morinaga, F.~Le~Kien, and K.~Hakuta,
  \enquote{{Measurement of fluorescence emission spectrum of few strongly
  driven atoms using an optical nanofiber},} Opt Express \textbf{18},
  17154--17164 (2010).

\bibitem{DeMartini:1996vh}
F.~De~Martini, G.~Di~Giuseppe, and M.~Marrocco, \enquote{{Single-mode
  generation of quantum photon states by excited single molecules in a
  microcavity trap},} Phys. Rev. Lett.  (1996).

\bibitem{Fleury:2000uf}
L.~Fleury, J.-M. Segura, G.~Zumofen, B.~Hecht, and U.~P. Wild,
  \enquote{{Nonclassical photon statistics in single-molecule fluorescence at
  room temperature},} Phys. Rev. Lett. \textbf{84}, 1148--1151 (2000).

\bibitem{Du:2008do}
S.~Du, P.~Kolchin, C.~Belthangady, G.~Y. Yin, and S.~E. Harris,
  \enquote{{Subnatural Linewidth Biphotons with Controllable Temporal Length},}
  Phys. Rev. Lett. \textbf{100}, 183603 (2008).

\bibitem{Reynaud:1983tq}
S.~Reynaud, \enquote{{La fluorescence de r{\'e}sonance: {\'e}tude par la
  m{\'e}thode de l'atome habill{\'e}},} Annales de physique  (1983).

\bibitem{Fleury:2001uw}
L.~Fleury, J.-M. Segura, G.~Zumofen, B.~Hecht, and U.~P. Wild, \enquote{{Photon
  statistics in single-molecule fluorescence at room temperature},} Journal of
  luminescence \textbf{94}, 805--809 (2001).

\bibitem{Loudor}
R.~Loudon, \enquote{{The quantum theory of light},}   (2000).

\bibitem{OKOSHI:1980un}
T.~Okoshi, K.~Kikuchi, and A.~Nakayama, \enquote{{Novel Method for
  High-Resolution Measurement of Laser Output Spectrum},} Electronics Letters
  \textbf{16}, 630--631 (1980).

\bibitem{Richter:1986us}
L.~E. Richter, H.~I. Mandelberg, and M.~S. Kruger, \enquote{{Linewidth
  determination from self-heterodyne measurements with subcoherence delay
  times},} Ieee J Quantum Elect \textbf{QE-22}, 2070 (1986).

\bibitem{Martienssen:1964ig}
W.~Martienssen, \enquote{{Coherence and Fluctuations in Light Beams},} American
  Journal of Physics \textbf{32}, 919 (1964).

\bibitem{Scarcelli:2004bc}
G.~Scarcelli, A.~Valencia, and Y.~Shih, \enquote{{Experimental study of the
  momentum correlation of a pseudothermal field in the photon-counting
  regime},} Phys. Rev. A \textbf{70}, 051802 (2004).

\bibitem{ESTES:1971}
L.~E. Estes, u.~Lorenzo, Narducci, and A.~T. Richard, \enquote{{Scattering of
  Light from a Rotating Ground Glass},} J. Opt. Soc. Am. \textbf{61},
  1301--1306 (1971).

\bibitem{Yuan:2009ue}
L.~Yuan, Z.~Yu-Chi, Z.~Peng-Fei, G.~Yan-Qiang, L.~Gang, W.~Jun-Min, and
  Z.~Tian-Cai, \enquote{{Experimental Study on Coherence Time of a Light Field
  with Single Photon Counting},} Chinese Physics Letters \textbf{26} (2009).

\end{thebibliography}
\end{document}